\documentclass[aps,twocolumn,preprintnumbers,amsmath,amssymb]{revtex4}
\usepackage{graphicx}
\usepackage{dcolumn}
\usepackage{bm}
\usepackage{multirow}
\usepackage{color}
\usepackage[colorlinks=true]{hyperref} 
\usepackage{amsmath}
\usepackage{amssymb}
\usepackage{epsfig}
\usepackage[T2A]{fontenc}
\usepackage[cp1251]{inputenc}
\usepackage[russian,english]{babel}
\usepackage{array}

\begin{document}

\title{Electrically tunable dynamic nuclear spin polarization in GaAs quantum dots \\ at zero magnetic field 
}

\author{M. Manca$^1$}
\author{G. Wang$^1$}
\author{T. Kuroda$^2$}
\author{S. Shree$^1$}
\author{A. Balocchi$^1$}
\author{P. Renucci$^1$}
\author{X. Marie$^1$}
\author{M. V. Durnev$^3$}
\author{M. M. Glazov$^3$}
\author{K. Sakoda$^2$}
\author{T. Mano$^2$}
\author{T. Amand$^1$}
\author{B. Urbaszek$^1$}
\email{urbaszek@insa-toulouse.fr}

\affiliation{$^1$Universit\'e de Toulouse, INSA-CNRS-UPS, LPCNO, 135 Avenue Rangueil, 31077 Toulouse, France}
\affiliation{$^2$National Institute for Material Science, Namiki 1-1, Tsukuba 305-0044, Japan}
\affiliation{$^3$Ioffe Institute, 194021 St.\,Petersburg, Russia}

\date{\today}

\begin{abstract}
In III-V semiconductor nano-structures the electron and nuclear spin dynamics are strongly coupled. Both spin systems can be controlled optically. The nuclear spin dynamics is widely studied, but little is known about the initialization mechanisms. Here we investigate optical pumping of carrier and nuclear spins in charge tunable GaAs dots grown on 111A substrates.  We demonstrate dynamic nuclear polarization (DNP) at zero magnetic field in a single quantum dot for the positively charged exciton X$^+$ state transition. We tune the DNP in both amplitude and sign by variation of an applied bias voltage V$_g$. Variation of $\Delta$V$_g$ of the order of 100~mV changes the Overhauser splitting (nuclear spin polarization) from -30~$\mu$eV (-22~\%) to +10~$\mu$eV (+7~\%), although the X$^+$ photoluminescence polarization does not change sign over this voltage range. This indicates that absorption in the structure and energy relaxation towards the X$^+$ ground state might provide favourable scenarios for efficient electron-nuclear spin flip-flops, generating DNP during the first tens of ps of the X$^+$ lifetime which is of the order of hundreds of ps. Voltage control of DNP is further confirmed in Hanle experiments.
\end{abstract}

\pacs{73.20.-r, 73.21.Fg, 73.63.Hs, 78.67.De}

                            \keywords{Quantum dots, optical selection rules}
\maketitle

\textit{Introduction.---} The spin coherence of carriers in semiconductor quantum dots (QDs) is limited by interactions with the nuclear spin bath fluctuations because the carrier
and nuclear spins are efficiently coupled through the hyperfine interaction \cite{Merkulov:2002a,Hanson:2007a,Khaetskii:2002a,Urbaszek:2013a,Bechtold:2015a,Kuhlmann:2013a,chekhovich2013element}, especially in III-V nano-structures with Ga, Al and In where 100~\% of nuclei have non-zero nuclear spin.
On the other hand, the stable nuclear spins themselves can potentially be used as a resource for quantum information storage \cite{Taylor:2003a,Morton:2008a,2017arXiv170204129A,dutt2007quantum,fuchs2011quantum}. Semiconductor quantum dots allow manipulating a mesoscopic ensemble of several thousand nuclear spins by optical manipulation of a single carrier spin, as both systems are efficiently coupled by the hyperfine interaction. 
Commonly nuclear spin polarization manipulation is achieved in applied magnetic fields \cite{Maletinsky:2009a,Greilich:2007a,Waeber:2016a,Nilsson:2013a}. In this work we show that both the sign and amplitude of optically generated dynamic nuclear polarization can be switched electrically in experiments at zero magnetic field. Here we use GaAs droplet dots in AlGaAs barriers which are a very versatile model system. They are ideally suited for quantum optics applications, with high entanglement fidelity for (111) grown dots \cite{Kuroda:2013a} and possible wavelength tuning to 780~nm to initialize rubidium (Rb) atoms for quantum memory experiments \cite{Basso:2017a,Akopian:2011a}. They strongly interact with nuclear spins \cite{Sallen:2014a,Vidal:2016a}, but one of the main interactions governing the nuclear spin dynamics, namely nuclear quadrupole effects due to strain \cite{Chekhovich:2015a,Sokolov:2016a}, is substantially reduced in this lattice matched system. 
Our experiments shed light on the importance of optical initialization of nuclear spins. In our charge tunable structure neither the main emission state (X$^+$) nor the sign of its photoluminescence (PL) polarization changes over a broad bias range of several hundred mV. But over this bias range we find that nuclear spin polarization can be tuned from -22~\%, going through zero to +7~\%. This indicates that the nuclei are polarized during optical absorption and subsequent relaxation towards X$^+$. We also present Hanle spin depolarization experiments for electrons and nuclei in transverse magnetic fields, which confirm the dependence of DNP on the applied bias.\\
\begin{figure*}
\includegraphics[width=0.98\linewidth]{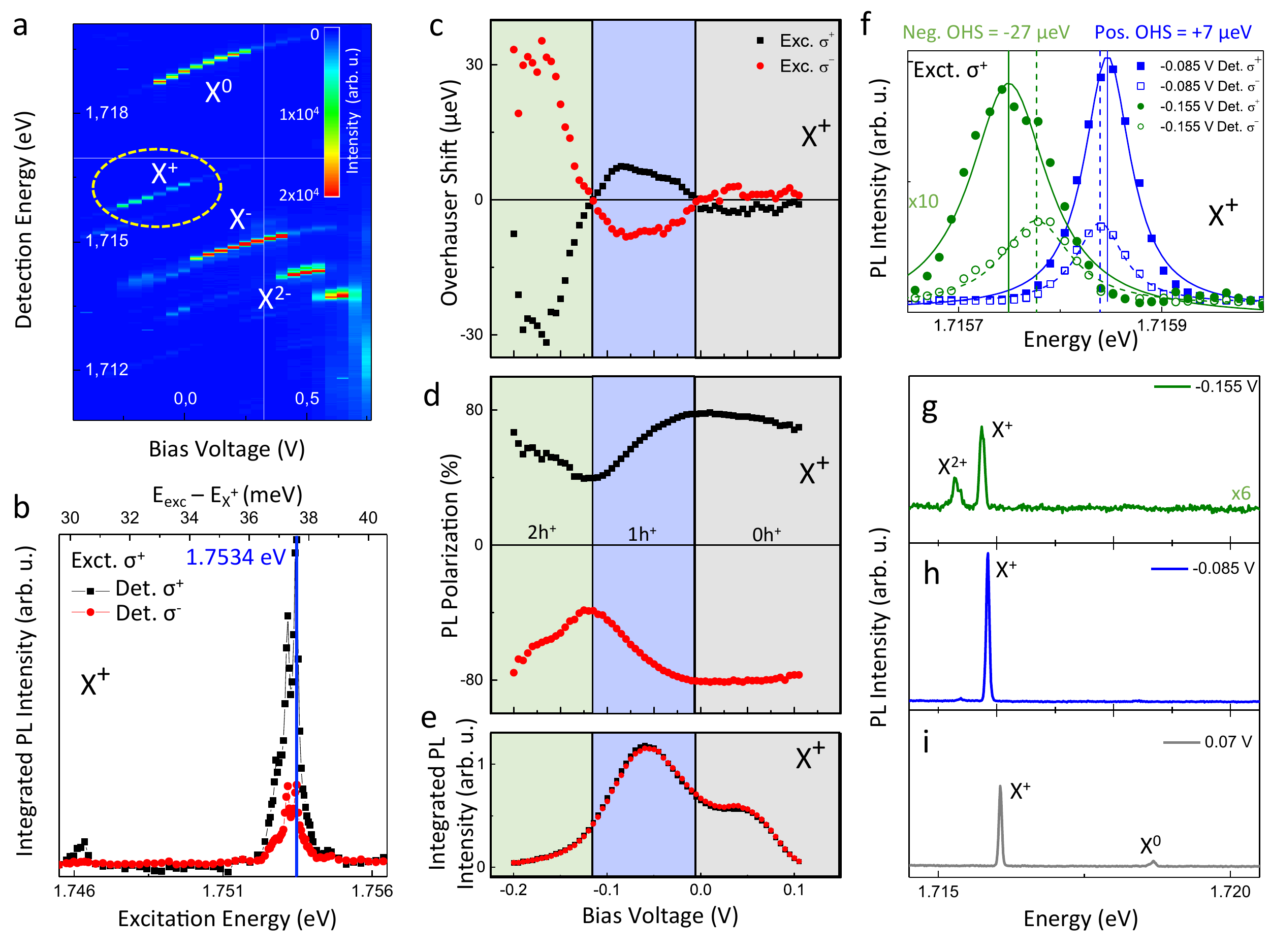}
\caption{\label{fig:fig1} (a) Contour plot of the single dot PL at $T=4$~K as a function of the applied bias voltage using a HeNe laser for above barrier excitation. The neutral exciton X$^0$ and charged excitons (trions) X$^+$ and X$^-$ are indicated. Blue means less than 100 counts, red $>15000$ counts for 30~s exposure time and 15~$\mu$W excitation power focussed on a 1$\mu$m diameter spot. (b) PL excitation for X$^+$ detection shows a resonance in intensity and polarization for a laser energy about 1 LO Phonon above the X$^+$ transition. Here the bias is V$_g$=-0.06~V. (c) Overhauser shift for X$^+$ transition as a function of applied bias. (d) Polarization of X$^+$ PL as a function of bias. (e) Intensity of X$^+$ PL as a function of bias. (f) Comparison of amplitude and sign of X$^+$ Overhauser shift OHS for bias values -0.085~V (blue symbols) and -0.155~V (green symbols). (g)-(i) Example PL spectra for initial occupation of dot with 2,1 and 0 holes, respectively. }
\end{figure*}
\begin{figure}
\includegraphics[width=0.98\linewidth]{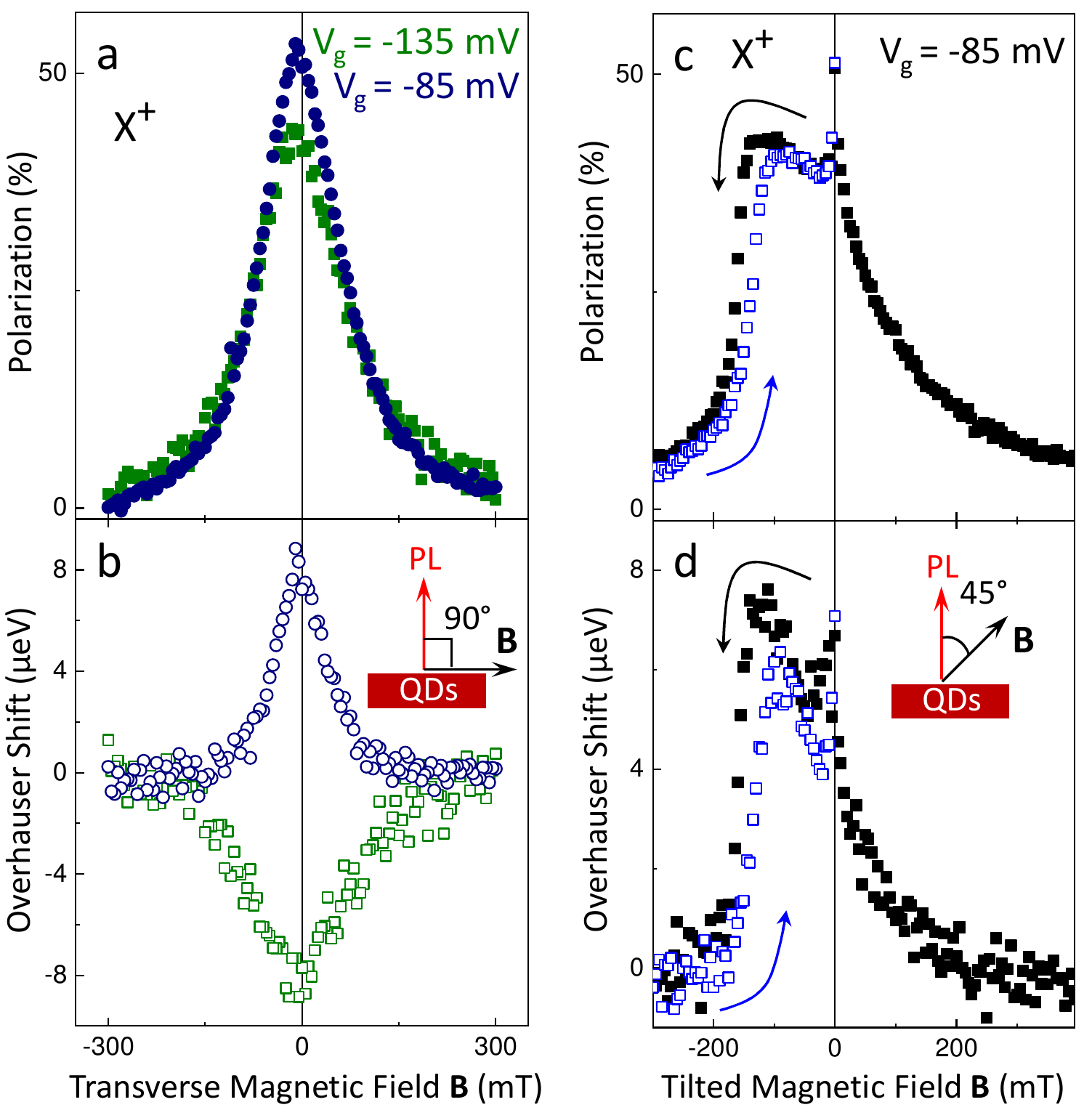}
\caption{\label{fig:fig2} (a) PL polarization of X$^+$ emission as a function of applied \textit{transverse magnetic field} for bias of $V_g=-0.135$~V (green squares) and $V_g=-0.085$ (blue circles). (b) Overhauser shift of X$^+$ emission as a function of applied \textit{transverse magnetic field} for bias of $V_g=-0.135$~V (green hollow squares) and $V_g=-0.085$ (blue hollow circles). (c) PL polarization of X$^+$ emission as a function of applied  magnetic field tilted 45 deg for bias of $V_g=-0.085$~V.  (d) Overhauser shift of X$^+$ emission.}
\end{figure} 
\textit{Samples and Experimental Set-up.---} The sample we study is the same as in Ref.~\onlinecite{Bouet:2014a}. It was grown by droplet epitaxy using a standard molecular beam epitaxy system \cite{Mano:2010a,Sallen:2011a,Belhadj:2008a,Belhadj:2010a}. 
and an $n^+$-GaAs(111)A substrate. The growth sequence is 50-nm $n$-GaAs (Si: $10^{18}$~cm$^{-3}$), 100-nm $n$-Al$_{0.3}$Ga$_{0.7}$As (Si: $10^{18}$~cm$^{-3}$), 20-nm Al$_{0.3}$Ga$_{0.7}$As tunnel barrier, GaAs QDs, 120-nm Al$_{0.3}$Ga$_{0.7}$As, 70-nm Al$_{0.5}$Ga$_{0.5}$As, and 10-nm GaAs cap.  Contrary to Stranski-Krastanov dots and dots formed by quantum well interface fluctuations \cite{Bracker:2005a} in our sample dots are not connected by a 2D wetting layer \cite{Mano:2010a,Sallen:2014a}. A 6~nm thick semitransparent Ti/Au layer serves as a Schottky top gate.\\
\indent The single dot PL is recorded at a temperature of 4~K with a home build confocal microscope with a detection spot diameter of $\simeq 1$~$\mu$m \cite{Durnev:2013a,Sallen:2014a}. The detected PL signal is dispersed by a double spectrometer and detected by a Si-CCD
camera with the spectral precision of 1~$\mu$eV. Nonresonant optical excitation used for initial dot characterization is achieved by pumping the AlGaAs barrier with a HeNe laser at 1.96~eV. For intra-dot quasi-resonant excitation typically 1~LO-phonon above the transition energy,  a tunable cw Ti-Sa laser Solstis from M Squared is used. The laser polarization control and PL polarization analysis are performed with Glan-Taylor polarizers and liquid crystal waveplates. Magnetic fields are generated by a vector magnet in an attoDry cryostat. \\
\indent \textit{Results and Discussion.---} In our charge tunable device, we can study independently positively charged trions X$^+$, neutral excitons X$^0$ and negatively charged trions X$^-$, see Fig.~\ref{fig:fig1}(a) and Refs.~\cite{Bouet:2014a,Vidal:2016a} for details. In reference \cite{Sallen:2014a} we reported and analysed detailed Hanle measurements of the X$^+$, but not in a charge tunable device, so here we can report in addition bias control of the spin interactions in [111] grown quantum dots. In our device, the carrier-nuclear spin interaction is probed through the  X$^+$ radiative recombination with the PL circular polarization $P_{c}=({I_{\sigma+}-I_{\sigma-})}/({I_{\sigma+}+I_{\sigma-})}$ directly proportional to the average spin $z$-component of the unpaired electron, $P_c = - 2 \langle S^e_z \rangle$. \\
\indent We perform PL excitation experiments (PLE) to identify efficient excitation energies for optical pumping. We find a PLE resonance about 37~meV above the X$^+$ emission energy in Fig.~\ref{fig:fig1}(b), which is most likely associated to phonon-assisted absorption, as the LO phonon energy in GaAs is $\approx36~$meV. Note that in PLE we find not just one peak but a doublet which might indicate that in addition to phonon assisted absorption other processes play a role, such as excitation of the first light hole level, which is estimated to be roughly 30~meV above the ground state transitions \cite{Durnev:2013a}. Transitions involving light holes can in principle generate electron spin polarization of opposite sign with respect to the transitions involving heavy holes in the quantum dot \cite{Dyakonov:2008a,Meier:1984a}. Tuning the excitation laser to this particular energy, allows the observation of very high PL polarization of $P_c=80\%$, indicating efficient carrier spin initialization in the dot. In a control experiment, keeping the same definition for $P_c$, we observe down to $P_c=-80\%$ when switching the laser excitation to $\sigma^-$, compare black with red data points in Fig.~\ref{fig:fig1}(d). \\
\indent Interestingly, pumping the dot quasi-resonantly with circularly polarized excitation results in a measurable splitting of the X$^+$ emission line into two lines in the absence of any applied magnetic field \cite{Lai:2006a,Dzhioev:2007a}, with one line polarized $\sigma^+$, the other $\sigma^-$. We observe an energy splitting $|E(\sigma^+)-E(\sigma^-)|$ up to 30~$\mu$eV, as shown in Fig.~\ref{fig:fig1}(f). This Overhauser splitting \cite{Overhauser:1953a} is a clear sign of dynamic nuclear polarization \cite{Gammon:1997a}, plotted in Fig.~\ref{fig:fig1}(c). As a control experiment, we show that the sign of the Overhauser shift is reversed as the laser polarization is switched from $\sigma^+$ to $\sigma^-$ in Fig.~\ref{fig:fig1}(c). \\
\indent As we measure the Overhauser shift of the X$^+$ transition as a function of bias we make a very surprising observation: 
the sign of the Overhauser shift and therefore of the dynamic nuclear polarization changes as a function of bias. 
We show example spectra in Fig.~\ref{fig:fig1}(f) for Overhauser shifts of $-27~\mu$eV and $+7~\mu$eV, respectively, corresponding to a nuclear spin polarization amplitude of 22\% and 5\%, respectively, in absolute value. 
In our experiment the sign of $P_c$ of the X$^+$ PL does not change for a given excitation polarization, see Fig.~\ref{fig:fig1}(d). 
This demonstrates that the final electron spin configuration is not indicative of the nuclear spin polarization, as expected for the flip-flop process enabled by the \textit{isotropic} or \textit{axially symmetric} hyperfine coupling. Indeed, in this situation the electron-nuclear flip-flop terms have the form:
\begin{equation}
\label{axial}
\mathcal H_{\rm flip-flop} \propto \hat I_+ \hat S_- + \hat I_- \hat S_+,
\end{equation}
where $\hat I_{\pm} = (\hat I_x \pm \mathrm i \hat I_y)/\sqrt{2}$ [$\hat S_{\pm} = (\hat S_x \pm \mathrm i \hat S_y)/\sqrt{2}$] are the nuclear [electron] spin rising (``+'') or lowering (``$-$'') operator and $x$, $y$ are the in-plane components. According to Eq.~\eqref{axial}, the electron spin angular moment is transferred to the nuclear spin systems and nuclear spins should align along the spin of the photoelectron. Thus, the nuclear spin polarization is generated possibly during absorption \cite{Bracker:2008a,Kloeffel:2011a,Latta:2009a,Chekhovich:2010a} and subsequent relaxation towards the X$^+$ ground state.\\
\indent In  Fig.~\ref{fig:fig1}(c) we identify 3 regions of bias, using $\sigma^+$ polarization: the Overhauser shift is negative for $V_g<-0.12~$V, then shows positive values and than becomes negative with very small amplitude for $V_g>0$. The entire bias range covering the 3 regions is dominated by X$^+$ emission. But although the light is emitted by the trion X$^+$, the dot can before photon absorption contain two, one or zero holes, as has been discussed in detail for the case of X$^-$ formation starting with two, one or zero electrons in the dot \cite{Laurent:2005a,Kloeffel:2011a,Simon:2011a}. It seems that this initial situation impacts the sign of the nuclear spin polarization. This would mean that the electron-nuclear spin flip-flop, necessary for dynamic nuclear polarization, occurred during formation of the X$^+$ configuration, with the 2 holes in a spin singlet. In Fig.~\ref{fig:fig1}(g-i) we show typical emission spectra for the regions with negative, positive and nearly zero Overhauser shift, respectively. In Fig.~\ref{fig:fig1}(g) we see in addition to the X$^+$ PL also a small additional peak, which we have identified as the X$^{2+}$ due to the very specific PL emission in applied magnetic fields \cite{Durnev:2016a}. In Fig.~\ref{fig:fig1}(h) the strong PL of the X$^+$ is the only visible transition, whereas for higher bias in Fig.~\ref{fig:fig1}(i) also the neutral exciton X$^0$ becomes visible.\\
\indent The results above can indicate that the non-collinear interaction of the charge carrier and nuclear spins, which is allowed in our trigonal [111]-grown quantum dots~\cite{Vidal:2016a} even in the absence of strain, can be of importance for the DNP~\cite{Hogele:2012a,Latta:2009a,xu09}. Symmetry analysis in Ref.~\cite{Vidal:2016a} demonstrates a variety of contributions beyond simple Eq.~\eqref{axial}. For example, due to the fact that the $C_{3v}$ symmetry of the quantum dot does not distinguish angular momentum components $\pm 1$ and $\mp 2$~\cite{Sallen:2011a} the following process becomes allowed: The electron spin initially oriented along the positive $z$-axis changes by $-1$, while the nucleus changes its spin by $-2$ (the total spin changes by $-3$). For the opposite helicity of excitation the electron spin can change by $+1$ giving rise to the nuclear spin flip by $+2$ (the total spin changes by $+3$). The contribution to the DNP due to such processes is opposite to that given by Eq.~\eqref{axial}. Additional non-collinear contributions are expected from the symmetry analysis for the heavy-hole-nucleus interaction~\cite{Vidal:2016a} which may also lead to opposite directions of electron and nuclear spins. The strengths of these contributions may depend, in particular, on the quantum dot charge state, e.g., due to the charge state dependence of the electric field gradients and nuclear quadrupole splittings. A detailed study of DNP scenarios and a quantitative model is beyond the scope of this work, but we interpret our surprising bias dependence of DNP in Fig.~\ref{fig:fig1}(c) as a fingerprint of the rich hyperfine coupling in 111 grown quantum dots.\\
\indent At first it might seem surprising that the bias region with the strongest X$^+$ PL does not correspond to the bias region with the strongest amplitude of the Overhauser shift. But energy conservation for DNP build-up is a major obstacle at zero applied \textit{magnetic} field, as the emergence of the Overhauser splitting will slow down the build up process which relies on electron-nuclear spin flip-flop. Here an uncertainty in energy, given by the correlation time $\tau_c$, is needed \cite{Dyakonov:2008a}. The correlation time of the electron-nuclear spin hyperfine interaction in a quantum dot can be extracted from magnetic field dependent measurements of the Zeeman splitting, as for example in Refs.~\cite{Maletinsky:2007b,Belhadj:2008a,Urbaszek:2013a}. From these experiments 
 $\tau_c$ of the order of a few tens of picoseconds can be extracted. This is shorter than the radiative lifetime of 0.5 to 1.0~ns in these systems and indicates that already in these experiments, events on shorter time scales during relaxation play a role. Also carrier tunnelling events during the radiative lifetime can play a role, as indicated in experiments at several Tesla for InGaAs quantum dots that showed a variation of $\tau_c$ with the applied electric field \cite{Nilsson:2013a}. 
Therefore DNP build-up in the experiments described here is likely to occur during absorption / relaxation processes, as energy conservation is relaxed by the short, fluctuating time $\tau_c$ which characterizes the electron - nuclear spin hyperfine interaction.
There the Overhauser shift of $-30~\mu$eV in the bias region $V_g<-0.125~$V might indicate a very favourable, short correlation time of the hyperfine interaction \cite{Urbaszek:2007a}. Note that absorption in this bias range can still be efficient, but the large electric field can dissociate the exciton complex as we approach the photocurrent regime, investigated in detail in charge tunable quantum dots \cite{Ramsay:2010a,Zrenner:2002a}. \\
\indent We aim to further study the coupling of carrier to nuclear spins in our dots in magnetic field dependent measurements. In Fig.~\ref{fig:fig2}(a) we compare $P_c$ as a function of the applied in-plane field for two different bias values. The polarization decays as in a typical Hanle measurement. For $V_g=-0.085~$V the Overhauser shift in Fig.~\ref{fig:fig2}(b) is negative and shows a very similar Hanle curve as the electron spin. This implies that over the investigated nuclear spin polarization range the nuclear spin polarization is proportional to the electron spin polarization. In contrast, for a bias value of $V_g=-0.085~$V the Overhauser shift is positive and decays also with a simple Lorentzian dependence \cite{foot1}. The data in Fig.~\ref{fig:fig2}(b) confirms clearly the sign of the observed DNP which switches with applied bias. \\
\indent A trademark for strong nuclear spin effects, especially in quantum dots, are bistability effects \cite{Urbaszek:2013a,Meier:1984a,Braun:2006a,Tartakovskii:2007a,Maletinsky:2007b,Kaji:2008a}. 
When we apply a tilted magnetic field, as in Fig.~\ref{fig:fig2}(c,d) the projection of the generated nuclear field onto the external field direction is enhanced, so compensation between the externally applied magnetic field and the nuclear field becomes possible \cite{Meier:1984a}. This compensation occurs after an initial drop of both the DNP and the electron spin polarization roughly at $B=-150$~mT, where the DNP recovers its zero field value. Here we observe a strong hysteresis, i.e. the measured electron spin and nuclear spin polarization for a given field depend on the sweep direction. This hysteresis can only be observed for the region where the external magnetic field is at least in part cancelled by the hyperfine field, in our case for negative applied magnetic fields.\\
\indent In conclusion, through the application of a small bias we can tune the dynamic nuclear polarization in an optically pumped quantum dot. Tuning between -22~\% to +7\% is achieved over a bias range of 100~mV without the need of external magnetic field for this change in sign and amplitude of the nuclear spin polarization. To further investigate bi-directional nuclear spin polarization reported here experiments under resonant excitation, referred to as "dragging" in the strained InGaAs dot system \cite{Hogele:2012a,Latta:2009a,xu09} could be carried out, to shed light on the non-collinear hyperfine interaction which goes beyond the simple description of DNP build-up via spin flip-flops. This non-collinear hyperfine interaction has been linked to the presence of strain i.e. strong nuclear quadrupole effects and it would be interesting to verify this hypothesis in our nearly strain-free sample system. \\
\indent We acknowledge funding from ERC Grant No. 306719, Marie Sklodowska-Curie actions ITN Spin-NANO Nr. 676108 and ITN 4PHOTON Nr. 721394 and LIA CNRS---Ioffe RAS ILNACS, RFBR Grant 17-52-16020. T.K. and T.M. thank Grant-in-Aid from JSPS.

 

\end{document}